\documentclass[twocolumn]{aastex62}

\usepackage{breakcites}
\hypersetup{breaklinks=true}

\usepackage{savesym}
\savesymbol{tablenum}
\usepackage{siunitx}	
\restoresymbol{SIX}{tablenum}
\usepackage{soul}
\graphicspath{{./}{figures/}}

\submitjournal{ApJ}

\shorttitle{Fragmentation in Pop III Galaxies}
\shortauthors{Kulkarni et al.}

\begin{document}

\title{Fragmentation in Population III Galaxies formed through Ionizing Radiation}

\correspondingauthor{Mihir Kulkarni}
\email{mihir@astro.columbia.edu}

\author[0000-0002-9789-6653]{Mihir Kulkarni}
\affiliation{Department of Astronomy, Columbia University, 550 West 120th Street, New York, NY, 10027, U.S.A.}

\author{Eli Visbal}
\affiliation{Center for Computational Astrophysics, Flatiron Institute, 162 5th Ave, New York, NY, 10010, U.S.A.}

\author[0000-0003-2630-9228]{Greg L. Bryan}
\affiliation{Department of Astronomy, Columbia University, 550 West 120th Street, New York, NY, 10027, U.S.A.}
\affiliation{Center for Computational Astrophysics, Flatiron Institute, 162 5th Ave, New York, NY, 10010, U.S.A.}



\begin{abstract}
Population III stars forming in minihalos tend to be relatively inefficient, with each minihalo hosting one or a small number of stars which are more massive than local stars but still challenging to observe directly at high redshift.  Here we explore a possible mechanisms for the generation of larger clusters of such stars: a nearby ionizing source which ionizes a late forming halo, delaying its collapse until the halo is sufficiently large that the core can self-shield and suffer runaway collapse.  We use simulations with a simple but accurate model for the radiative ionizing flux and confirm the basic predictions of previous work: higher ionizing fluxes can delay the collapse to lower redshifts and higher masses, up to an order of magnitude above the atomic cooling limit.  In a limited number of runs we also examine the fragmentation of the cores at even higher resolution, using both simple estimates and sink particles to show that the number of fragments is generally small, at most a handful, and that the mass accretion rate on the fragments is of order $10^{-3}$ M$_\odot$ yr$^{-1}$.  This rate is sufficiently high that the descent on the main sequence (and hence the suppression of accretion) is delayed until the stellar masses are of order $100-1000$ M$_\odot$, but not high enough to produce direct collapse black holes of mass $\sim 10^5$ M$_{\odot}$.  The resulting clusters are larger than those produced in minihalos, but are still likely to fall short of being easily detectable in JWST blind fields.
\end{abstract}

\keywords{stars: Population III -- galaxies: high-redshift -- cosmology: theory}




\section{Introduction}
Population III (Pop III) stars are the first generation of stars forming from primordial gas.
Numerical simulations of $\Lambda$CDM predict that Pop III stars typically form in minihalos of mass $10^5-10^6 \textrm{M}_\odot$ \citep{Haiman96, Tegmark97, Machacek, Abel2002, Bromm2002, Yoshida2003, Greif15}.
The typical virial temperature of these minihalos is below the atomic cooling threshold, so the gas cools via rotational-vibrational transition lines of $\text{H}_2$ in the absence of metals. 

Lyman-Werner (LW) radiation, in the range 11.2-13.6 eV, can photo-dissociate molecular hydrogen. 
As the star formation density in the Universe increases, it creates a LW radiation background that can suppress Pop III star formation in small mihihalos \citep{Haiman97, Machacek, Wise-Abel07,Oshea07, Wolcott11, Visbal14}. 
In regions with strong LW background radiation, Pop III star formation is suppressed in halos with virial temperatures up to \SI{e4}{K}, where atomic hydrogen cooling becomes efficient. 
This temperature corresponds to a mass of $\textrm{M}_{\textrm{vir}} \approx 3 \times 10^7 ((1+z)/11)^{-3/2} \si{ M}_{\odot}$ \citep{Barkana01}. 

The lifetime of massive Pop III stars is short, a few Myr \citep{Schaerer02}, which
leads to the enrichment of the gas with metals via supernova winds and the transition to the formation of lower-mass Pop II stars \citep{Bromm03_lowmass, Schneider06, Smith2007, Maio2010}.
This means that Pop III stars are typically not expected to be found in more massive halos because the metal cooling will lead to lower temperatures (and lower mass stars).
Although, inefficient metal mixing may lead to Pop III star formation at much lower redshift \citep{Scannapieco03, Jimenez2006}.

It has been suggested that, in the presence of a strong ionizing radiation source, Pop III stars can form in much more massive halos \citep{Johnson10, Visbal16_cr7, Visbal17, Yajima_Khochfar2017, Johnson18}. 
Previous works have investigated photoevaporation of minihalos because of ionizing radiation \citep{Shapiro04, Iliev05} and the suppression of star formation in more massive halos because of reionization \citep{Shapiro94, Thoul_Weinberg1996, Gnedin2000, Dijkstra04, Hoeft2006, Okamoto2008, Sobacchi2013, Noh2014}.

\cite{Visbal17} studied Pop III star formation in three halos subjected to various background ionizing fluxes. 
They concluded that, for high fluxes, ionization and photo-heating of the gas can delay cooling and collapse into stars in halos up to masses significantly higher than the atomic cooling threshold. 
The threshold halo mass for collapse increases with increasing flux, and saturates at a value approximately an order of magnitude above the atomic cooling threshold for very high fluxes. 
The ionizing flux keeps gas ionized and hot ($>$ \SI{e4}{K}) until the dark matter halo grows and the gas is compressed to sufficiently high density that self-shielding becomes important. 
This leads to a dense self-shielded core with an increased fraction of neutral hydrogen, causing the gas to cool and collapse.  

In this work, we further study the properties of such massive Pop III galaxies, focusing, in particular, 
on the fragmentation processes within the central core and the clump mass function in such a galaxy. 
We present our results for three halos that we have selected to study using zoom-in simulations, exploring their collapse properties, gas fragmentation and discussing the impact of ionizing radiation from the stars that form.  

This paper is structured as follows. In section \ref{sec:method}, we describe the set up for our cosmological zoom-in simulations, and how we have included background ionizing radiation. 
We also detail the sink particle adaption in \textsc{Enzo} that we have used to study the fragmentation. 
In section \ref{sec:results}, we describe the properties of collapse as a function of background ionizing flux, radial profiles in halos at the time of runaway collapse and properties of the sink particles such as their masses and accretion rates, and estimates of their corresponding stellar masses. 
In section \ref{sec:discussion}, we discuss the collapse criterion for a halo with a given background ionization flux in more detail, the effect of radiation feedback from the stars formed, as well as the expected frequency of our clusters and observational prospects with JWST. 
We summarize our conclusions in section \ref{sec:conclusion}.
\vspace{0.5cm}

\section{Methodology}
\label{sec:method}

\subsection{Simulation set-up}

We study the formation of Pop III stars in these massive halos with simulations using the adaptive mesh refinement (AMR) code \textsc{Enzo} \citep{Enzo}. It is an Eulerian mesh-based hydrodynamic code that subdivides cubic cells into 8 smaller cells when certain refinement criteria are met. This allows the code to resolve regions of interest, while not wasting computation time on the large volumes outside of collapsed halos. \textsc{Enzo} includes various gas species - atomic hydrogen, ionized hydrogen, molecular hydrogen, helium etc. It also has multiple hydro solvers; we use the energy conserving, spatially third-order accurate Piecewise Parabolic Method (PPM) here. The code follows the non-equilibrium evolution of 9 species (H, H$^+$, He, He$^+$, He$^{++}$, e$^-$, H$_2$, H$_2^+$ and H$^-$) and includes radiative processes through the Grackle library \citep{Grackle}. As the gas temperature in our simulations does not fall below \SI{150}{K}, we do not expect HD cooling to play a significant role \citep{McGreer08} and hence do not include it.

For this work, we run zoom-in cosmological simulations using a modified version of \textsc{Enzo} 2.5. We assume a $\Lambda$CDM cosmology consistent with \cite{Planck14} throughout: $\Omega_m = 0.32$, $\Omega_\Lambda = 0.68$, $\Omega_b = 0.049$, $h = 0.67$, $\sigma_8 = 0.83$, and $n_s = 0.96$. We generate the initial conditions for our simulations using \textsc{MUSIC} \citep{Music}. To identify halos we run 3 sets of zoom-in simulations centered around 3 different halos from a cosmological box of comoving size \SI{2}{h^{-1} Mpc}. We first run a dark-matter only simulation starting at redshift $z = 200$. We identify a halo in a redshift $z = 6$ snapshot and rerun the simulation with hydrodynamics using a refined region around the halo. The halos were selected such that they would be the most massive halo in the refined region, as we want the halo of interest to collapse first. We also make sure that the halo considered here is relatively far away from other similar mass halos. This is to avoid complications from tidal forces of neighboring massive halos. This may seem contradictory to our problem where we need a nearby Pop II star galaxy as a source of ionizing radiation; however, for simplicity, we wish to focus only on the collapsing halo in this work.  Future work should aim for a more self-consistent simulation. Halo A was selected from a $128^3$ dark matter only simulation with a virial mass of \SI{1.3e9}{M_{\odot}} at z = 6. We rerun the hydro simulation with a $128^3$ grid with 3 added (initial) refinement levels in the zoomed region, giving an effective initial resolution of $1024^3$ in the zoomed region. For halos B and C, we select them in a $256^3$ dark matter only simulation, so as to be able to select lower mass halos. Halos B and C have virial masses of \SI{1.22e8}{M_{\odot}} and \SI{3.2e8}{M_{\odot}}. We then rerun hydro simulations zoomed around these halos with a $128^3$ grid and 3 added initial refinement levels in the zoomed region. We choose the random seeds for the MUSIC initial condition generator such that the initial conditions are the same for the $256^3$ grid dark matter only and $128^3$ grid hydro runs. The dark matter particle mass was \SI{836}{M_{\odot}} in the zoomed region for all the simulation runs.

We refine the cells in the zoomed region into smaller cells based on the following criteria: baryon mass, dark matter mass and the Jeans length. We refine the cell if the baryon or dark matter density becomes higher than $4\times 2^{3 l}$ times the corresponding densities on the root grid i.e. the coarsest grid ($128^3$) in the simulation and $l$ is the refinement level, meaning that cells with more mass than 4 times the initial dark matter particle mass or baryon equivalent will be refined. The Jeans length is resolved by at least 16 cells, and generally controls the refinement during the later parts of the collapse.  Cells are refined down to 18 levels from the root grid, resulting in a best resolution of 0.085 pc comoving.
\subsection{Ionization implementation}

We treat hydrogen ionization radiation in the same way as \cite{Visbal17}. This treatment is based on \cite{Rahmati}, which provided fitting functions to the photoionization rate as a function of local density that match cosmological simulations of the post-reionization universe with radiative transfer. We assume a uniform photoionzation rate in the box which is modified locally based on the self-shielding factor.

The hydrogen photoionization rate is given as
\begin{equation}
\Gamma = f_{\text{sh}} \Gamma_\text{bg},
\label{eq:local_gamma}
\end{equation}   
where $f_{\text{sh}}$ is the local self-shielding factor and $\Gamma_\text{bg}$ is the uniform ionizing background without shielding. The self-shielding factor depends on the temperature and density of hydrogen as
\begin{equation}
f_{\text{sh}} = 0.98 \left[1+\left(\frac{n_\text{H}}{n_\text{H,sh}}\right)^{1.64}\right]^{-2.28} + 0.02 \left[1+\frac{n_\text{H}}{n_\text{H,sh}}\right]^{-0.84},
\end{equation}
where
\begin{equation}
n_\text{H,sh} = \SI{5e-3}{cm^{-3}} \left(\frac{T}{10^4 \si{ K}} \right)^{0.17} \left(\frac{\Gamma_\text{bg}}{10^{-12} \si{s^{-1}}}\right)^{2/3}.
\end{equation}
We apply the same shielding factor to the hydrogen photo-heating rate as well. We assume a $T_* = 3 \times 10^4 \si{ K}$ black-body spectrum for computing the relative ionization and heating rates for hydrogen.

We use an ionizing flux in units of $F_0 = 6.7 \times 10^6$ photons s$^{-1}$ cm$^{-2}$, which is equivalent to having a source emitting $2 \times 10^{54}$ ionizing photons isotropically at a distance of $50$ kpc. According to merger-tree calculations in \cite{Visbal16_cr7}, a dark matter halo of mass $M = \SI{6.6e11}{ M}_{\odot}$ at $z \sim 7$, with star formation efficiency and escape fraction 0.1, will produce $2 \times 10^{53}$ photons per second over a range of redshift $\Delta z \approx 10$, which would correspond to flux of $0.1 F_0$ if the source were at a distance of 50 kpc. We do not self-consistently model this source within the box; as discussed later in this paper, and in more detail in \citep{Visbal16_cr7}, the detailed setup is quite rare and would require a much larger cosmological volume to model self-consistently.  In particular, by doing this, we assume that the only impact of the source is radiative.

Ideally, one should ray-trace the photons from the ionizing source to correctly simulate the problem. \cite{Visbal17} show that the results do not vary significantly if we use an ionization treatment based on \cite{Rahmati} instead of ray-tracing photons from an actual source. It is however computationally far less expensive. Hence, we use a uniform and isotropic radiation field (corrected for local extinction) in this work. Some unintended effects of this will be discussed below.

\cite{Visbal17} showed that the redshift at which ionizing radiation is turned on does not have a large effect on the results, as long as there is enough time for the intergalactic medium to be ionized completely. We start the background ionizing radiation at redshift $z = 30$. 

We include a uniform LW background radiation with flux $J_\text{LW}  = 100 \times J_{21}$ such that the H$_2$ photo-dissociation rate $k_{\text{H}_2} = 1.42 \times 10^{-12} J_\text{LW}/J_{21} \si{ s^{-1}}$. $J_{21}$ here is $10^{-21} \si{erg s^{-1} cm^{-2} Hz^{-1} Sr^{-1}}$. When ionizing radiation is present, we increase the LW radiation assuming a black-body spectrum corresponding to a temperature of 30,000 K. For ionizing flux $F_0$, the total LW background flux is $J_\text{LW}  = 175 \times J_{21}$. We modify the photo-dissociation rate by the self-shielding function described in \cite{Wolcott11}.

Our choice of background intensity $J_{\rm LW} = 100 J_{21}$, represents what could be found in an overdense region of the Universe \citep[see e.g. Figure~11 in][]{Ahn2009}. When ionizing radiation from a nearby galaxy is present, we increase the LW radiation as $J_{\rm LW} = (100 + 75 \times F/F_0) J_{21}$, where $F$ is the ionizing flux. This value of the flux assumes the source is a 30,000 K blackbody with an escape fraction in ionizing photons of $f_{\rm esc} \approx 0.2$

\subsection{Fragment mass and sink particle implementation}
\label{subsec:sinks_setup}

We define runaway collapse as the time when the simulation reaches the highest refinement level 18. Continuing the simulation further without refining cells leads to artificial fragmentation \citep{Truelove97}. To continue the simulation after the runaway collapse, we first used the \textit{MinimumPressureSupport} parameter in \textsc{Enzo} \citep{Machacek}. This parameter adds artificial pressure to the most refined cell so that the Jeans length of the most refined cell is resolved by the square root of the \textit{MinimumPressureSupport} number of cells. We set the \textit{MinimumPressureSupport} to 256 to make it consistent with the Jeans refinement parameter of 16 used. We then used YT's \citep{yt} clump finder to identify clumps at various times. While useful, the clump finder in YT does not connect clumps across snapshots and so returns different locations and properties for clumps in snapshots at different times. Hence, we cannot use the information from the clump finder to accurately find out how particular clumps are evolving in time. 

Therefore, in addition to the above estimates, we also used a simple sink particle technique to determine the accretion rates and clump mass evolution. Sink particles have been previously used for simulating first stars in SPH codes \citep[e.g.,][]{Stacy10, Stacy12, Greif12} and grid-based codes \citep[e.g.,][]{Krumholz2004}.
The evolution of the protostars depends on the accretion rates onto them. To estimate accretion rates this way, we reran the simulation and used a sink particle implementation in \textsc{Enzo} (instead of the \textit{MinimumPressureSupport} technique described above).  When reaching the highest refinement level, we add a sink particle to the most refined cell. This sink particle accretes mass from the cells around it such that the maximum density of any cell in a sphere of radius 5 cell widths is no higher than the density which would exceed the refinement criterion on that (maximum) level of refinement.  This insures that the minimum density is accreted such that the Jeans length criterion is not violated, avoiding artificial fragmentation.
By comparing the mass of the sink particles at various snapshots, we can calculate their accretion rates, which can be used to estimate the final stellar masses.  
We merge two sink particles into one if the distance between them is less than 10 times the width of the smallest cell. The gravitational force on sink particles is calculated at the highest refinement level (i.e. it is not smoothed).  
More sophisticated sink particle algorithms are possible \citep[e.g.,][]{Regan18}, and recent work \citep{Regan18b} has explored the fragmentation of atomic cooling halos exposed to a LW background (without the ionizing background adopted here).

\section{Results}
\label{sec:results}
\subsection{Collapse properties}
\label{subsec:collapse_properties}

Collapse is delayed to lower and lower redshifts for increasing background flux values. The collapse occurs when the cooling time and the dynamical time rapidly decrease and the heating time rapidly increases, marking the runaway collapse.  

\begin{figure}
\begin{center}
\includegraphics[width=\linewidth]{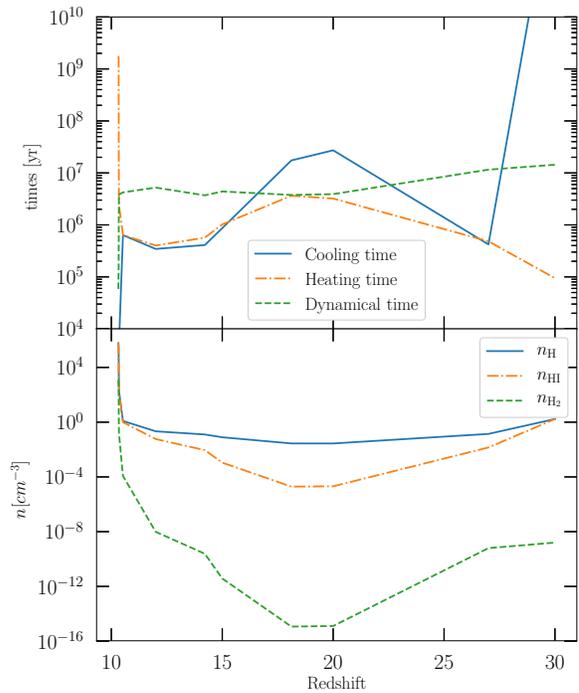}
\caption{Important timescales and densities in the central region of halo A for background flux $F_0$ as a function of redshift. The top panel shows the cooling, heating and dynamical times in the center of the halo as a function of redshift. The bottom panel shows the densities of the total, neutral and molecular hydrogen densities in the same region. At the point of runaway collapse, the densities increase rapidly which coincides with rapidly falling dynamical and cooling times, and a rapidly increasing heating time.}
\label{fig:collapse_times_densities}
\end{center}
\end{figure}

We can understand the physics of the collapse by looking at Figure~\ref{fig:collapse_times_densities}. The top panel shows the cooling, heating and dynamical times for the central region of halo A as a function of redshift for a background ionizing flux equal to $F_0$ and the bottom panel shows H density, \textsc{H i} density and H$_2$ density as a function of the redshift for the same halo. The background ionizing radiation is turned on at redshift 30, which starts the ionization and heating of the gas. This also decreases the cooling time significantly, as the atomic cooling is most efficient near the temperature of \SI{e4}{K}. With time, more gas in the central region gets ionized and we can see that the ionization reaches a peak near redshift $z=18$. As the potential well of the dark matter halo gets deeper and gas density increases, the gas is able to self-shield itself better from the background radiation. The runaway collapse occurs when the dynamical time and the cooling time become much shorter and cross the heating time which increases rapidly. 

As shown in \cite{Visbal17}, the collapse of the halo is delayed further with increasing background ionizing flux. The delays in redshift for a given change of background flux are different for different halos. In particular, halo B does not collapse untill redshift $z=5$ for background flux higher than $0.01 F_0$. This issue is discussed more in section~\ref{sec:discussion}.

\subsection{Radial profiles}

\begin{figure*}
\centering
\includegraphics[width=0.49\linewidth]{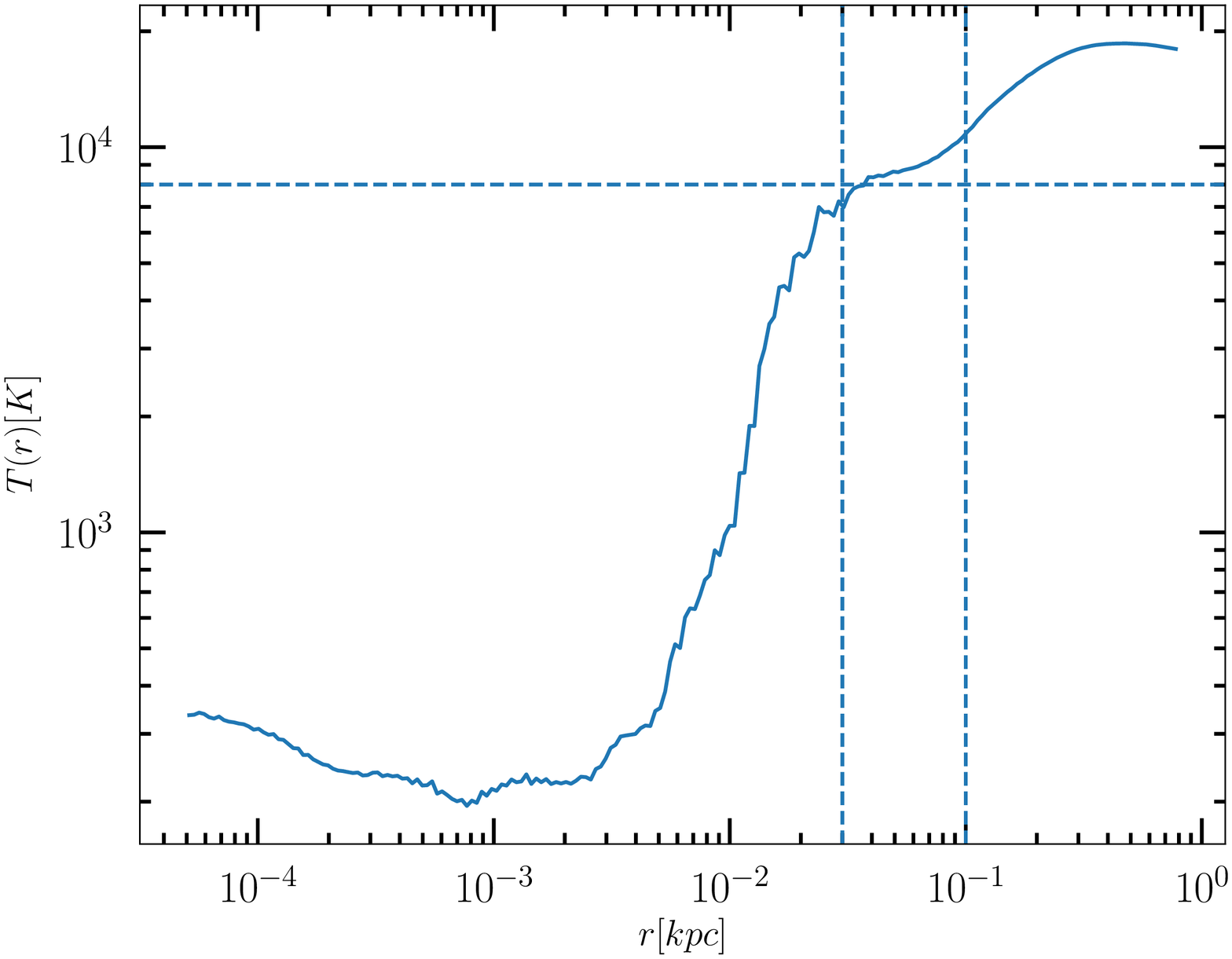}
\hfill
\includegraphics[width=0.49\linewidth]{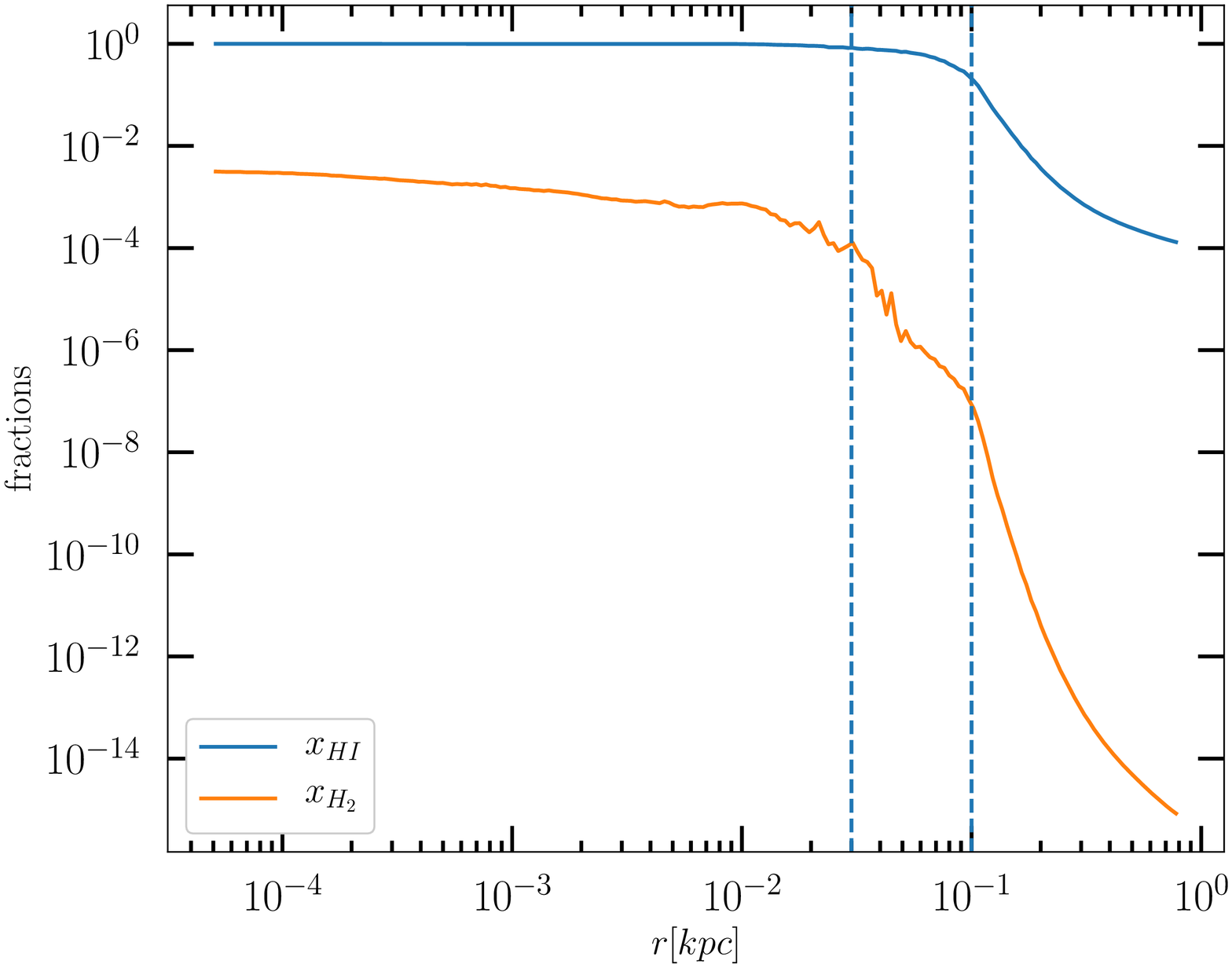}
\caption{Radial profiles of the gas temperature (left) and the neutral and molecular hydrogen fractions (right) for halo A at the point of runaway collapse with a background flux $F_0$ at $z = 10.33$. In the right panel, we can see that the gas forms a core of molecular hydrogen of radius $\sim 30$ pc and a neutral core of radius $\sim 100$ pc (shown with vertical blue dashed lines in both plots). This is consistent with the temperature profile in the left panel where $T < \SI{e4}{K}$ for $r < \SI{30}{pc}$ because of the molecular hydrogen cooling, $T \sim \SI{e4}{K}$ for $\SI{30}{pc} < r < \SI{100}{pc}$ because of atomic cooling and $T > \SI{e4}{K}$ for $r > \SI{100}{pc}$ because of the ionizing radiation. The horizontal blue dashed line in the temperature profile corresponds to a temperature of \SI{8e3}{K}.}
\label{fig:radial_profiles}
\end{figure*}

The radial gas density profile at the runaway collapse follows an $r^{-2}$ profile consistent with an isothermal profile \citep{Larson69} for all halos. 
The denser gas in the center can self-shield itself from the background ionizing flux and forms a neutral core. In our simulations, this is reflected through the lower value of the local $f_{sh}$ factor at high densities which reduces the effective ionizing radiation strength as shown in Eq.~\ref{eq:local_gamma}. 
Figure~\ref{fig:radial_profiles} shows the temperature as a function of radius in the left panel and the averaged neutral and molecular hydrogen fractions as a function of radius in the right panel, for halo A at the point of runaway collapse for a background flux $F_0$ at $z = 10.33$. 
We see a neutral core of size $\sim 100$ pc and also a smaller core of $\sim 30$ pc with high molecular hydrogen fraction. 
These length scales can also be traced in the temperature profile on the left.
The temperature of the ionized gas outside $100$ pc is higher than $10^4$ K because of the background ionizing radiation. 
In the region from $r = 30-100$ pc, the gas cools via H electronic transitions and hence has a temperature close to $10^4$ K.
The gas within 30 pc from the center is cooled mostly via rotational-vibrational transition lines of molecular hydrogen and hence the temperature falls from $10^4$ K to a few hundred Kelvin.
In all of our halos, we see similar temperature profiles which include ionized gas in the outer parts of the halo and cold gas in the self-shielded cores in the centers. 

\subsection{Sink/clump accretion}
\label{subsec:clump accretion}
In this subsection, we explain the properties of the sinks formed and their accretion histories.

The left panel of Figure~\ref{fig:haloA_projections} shows the projected density along x-axis for halo A at runaway collapse with background flux $F_0$ at $z = 10.33$. The contours show the clumps identified by YT's clump finder. 
The mass of the central dense clump grows as $\SI{e-3}{M_{\odot} yr^{-1}}$. We also calculate accretion on the sphere of radius 1 pc centered around the densest point in the halo as the rate of change of the enclosed mass which is also consistent with the accretion rate of $\SI{e-3}{M_{\odot} yr^{-1}}$.

Sink particles give us a better handle on the accretion rates on protostars not just at the center, but also forming at other locations as explained in the Section~\ref{subsec:sinks_setup}. 

The accretion rate on to the sink particles are consistent with the expected rate based on the gas temperature, given by $\dot{M} = c_s^3/G$, where $c_s$ is the sound speed and $G$ is the Gravitational constant.  This low accretion rate may initially be surprising given that the halo mass is quite large; however, we point out that, due to the self-shielding core, the collapse occurs in what is effectively a low-radiation environment, and the gas cools down to $\sim 500$ K.  The resulting density and temperature structure of the fragmenting clumps appear to be very similar to that produced by the collapse of minihalos.  Indeed, that the accretion rates we determine are roughly consistent with those in \citet{Regan18}, for similar LW backgrounds.

\subsubsection{Halo A}
\begin{figure*}
    \centering
    \includegraphics[width=\linewidth]{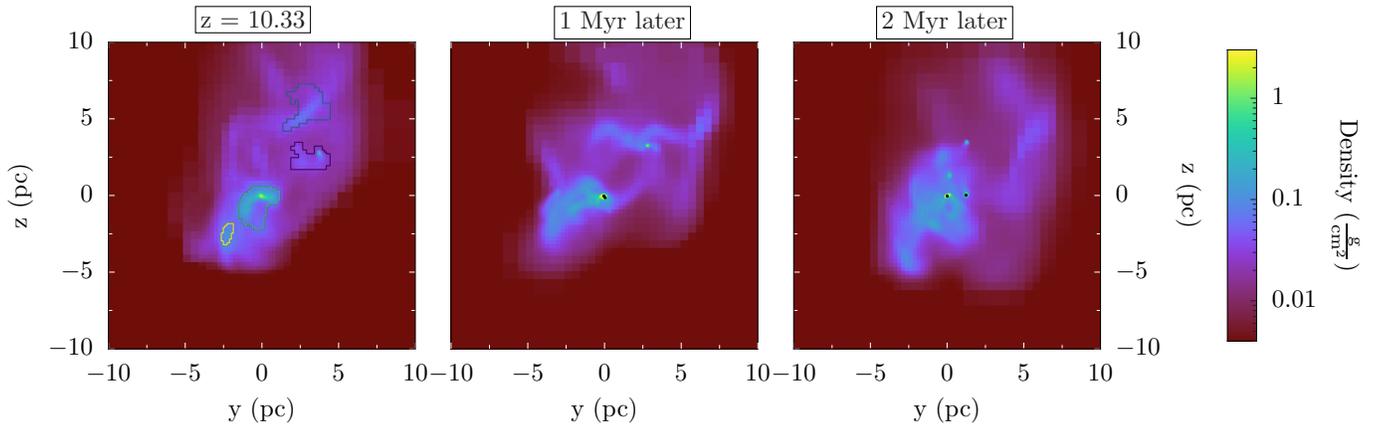}
    \caption{Projected densities along x-axis for halo A for the background flux $F_0$ at the runaway collapse at z = 10.33 (left panel), 1 Myr after the runaway collapse (middle) and 2 Myr after the runaway collapse (right). In the left panel, the contours show the clumps identified by YT's clump finder. The middle and right panels show the location of sink particles formed with black dots.}
    \label{fig:haloA_projections}
\end{figure*}
We evolve our simulation after the runaway collapse for the background flux of the $F_0$ case. Even though YT's clump finder identifies 4 clumps at the runaway collapse, not all of them form sink particles at the same time as they have not reached the highest refinement level. 
At runaway collapse, a sink particle is formed at the densest point in the halo which is at the center in the left panel of Figure~\ref{fig:haloA_projections}.
The middle and the right panels of Figure \ref{fig:haloA_projections} show projected densities at 1 Myr and 2 Myr after the runaway collapse, respectively.
Black dots denote the location of the sink particle formed. Nearly 2 Myr after the start of runaway collapse, another clump becomes dense enough to form a sink particle, which can be seen in the right panel of Figure~\ref{fig:haloA_projections}.

The mass of the most massive sink particle and its accretion rate as function of time is shown in the left panels of Figures \ref{fig:sinks_all}.

\subsubsection{Halo B}
The gas in halo B does not collapse for any background flux equal to or higher than $0.03 F_0$. The reasons for this happens are discussed in the discussion section~\ref{subsec:collapse criterion}. We select the halo with the highest background ionizing flux for which gas still undergoes runaway collapse ($F = 0.01 F_0$ in this case) to continue its evolution further in time. We add a sink particle in the densest cell at the point of runaway collapse. The total mass and accretion rate of that sink particle are shown in the middle panels of Figure~\ref{fig:sinks_all}.  

\subsubsection{Halo C}
For halo C, we chose the case with $F = 0.1 F_0$ to evolve further after runaway collapse. The runaway collapse occurs at $z = 6.54$. This is the halo in our simulations that shows the most fragmentation. We see multiple clumps forming with possible sites for the formation of sink particles. The left, middle and right panel of Figure~\ref{fig:haloC_projections} show projected densities at the point of runaway collapse, and 1 and 2 Myr after the runaway collapse, respectively. In the middle panel, we see 3 sink particles in the densest clumps. In the right panel, we see that some of the sink particles have drifted a bit from the clump in which they formed. 

\begin{figure*}
    \centering
    \includegraphics[width=\linewidth]{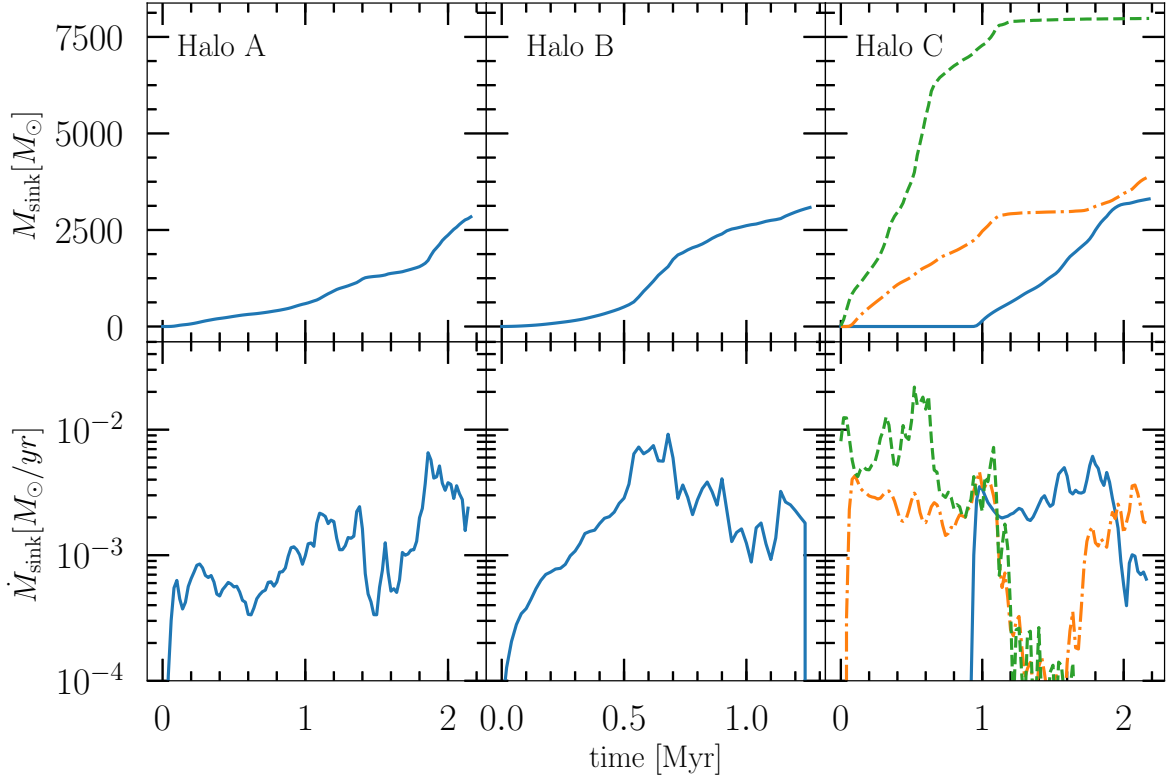}
    \caption{The left, central and right panels show the evolution of sink particles for the halos A, B and C respectively. The top panels show the total mass accreted by the sink particles as a function of time. For halo C, it shows the accretion history of the 3 most massive sink particles in the simulation. In the absence of proto-stellar feedback, the mass of the sink particle does represent the mass of the final star formed. The corresponding stellar masses can be estimated using the accretion rate of the sink particle (shown in the bottom panels) and a fitting function given by \cite{Hirano14}. Based on the fit, we estimate the stellar masses for the sinks shown to be $\sim \SI{40}{M_\odot}$ for halo A, $\sim \SI{100}{M_\odot}$ for halo B and 600, 100, 100 $\textrm{M}_\odot$ for halo C. As all the sink particles have accretion rates lower than $\dot{\textrm{M}}_\text{crit} = \SI{0.04}{M_\odot}$ \citep{Hirano14}, we do not expect any of the sinks to form high mass ($M > 10^{4}$ M$_\odot$ black hole seeds.}
    \label{fig:sinks_all}
\end{figure*}

\begin{figure*}
    \centering
    \includegraphics[width=\linewidth]{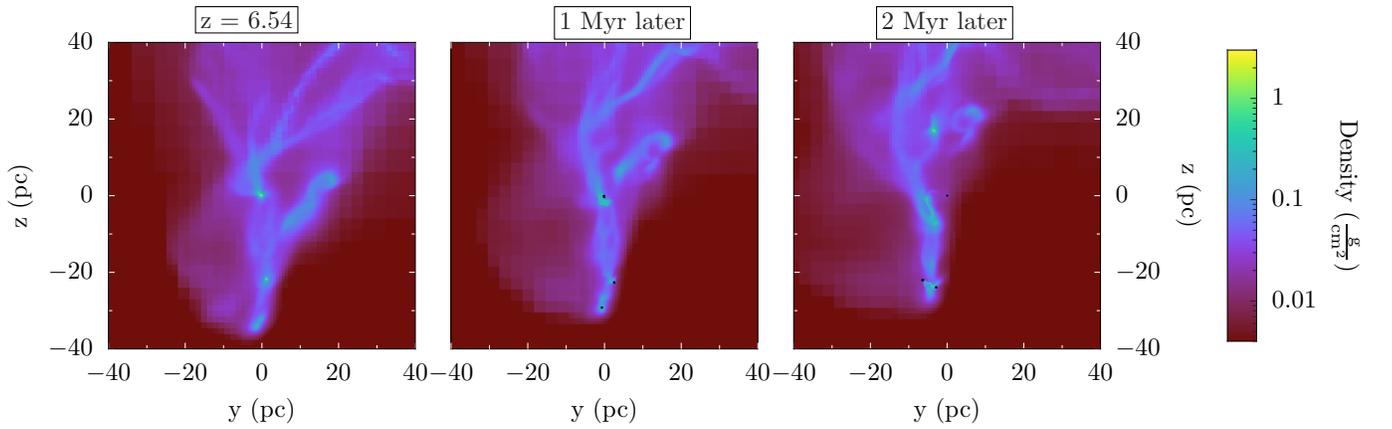}
    \caption{Projected densities along the x-axis for halo C for the background flux $0.1 F_0$ at the runaway collapse at z = 6.54 (left panel), 1 Myr after the runaway collapse (middle) and 2 Myr after the runaway collapse (right). The black dots in the middle and right panels show the locations of sink particles formed. Accretion histories of the sink particle are shown in Figure~\ref{fig:sinks_all}.}
    \label{fig:haloC_projections}
\end{figure*}

The accretion histories of the three most massive sink particles are shown in the right panels of Figure~\ref{fig:sinks_all}. The most massive sink particle accretes with a rate as high as $\SI{0.02}{M_\odot {yr}^{-1}}$. 

\subsection{Stellar mass estimates}
In this subsection, we describe how sink particles are used to estimate the corresponding mass of Pop III stars. Sink particles accrete mass from cells around them and grow in mass. The sink particle routine does not include any radiative feedback from the protostar that would lower the accretion rate and the final mass of the Pop III star formed. 

\cite{Hirano14} studied the evolution of Pop III protostars and their final masses including the relevant stellar physics for a sample of 100 first stars. 
Their sample is about Pop III star formation in minihalos that mostly form individual stars, a different environment from one we are studying here, however we expect that with similar accretion rates, the feedback impact for gas close to the star would not change. 
From their large sample, they find a linear relation between accretion rate on the protostar and the final mass of the Pop III star given as

\begin{equation}
M_\text{PopIII} = 100 M_{\odot} \left(\frac{\dot{M}}{2.8 \times 10^{-3} M_{\odot} \text{yr}^{-1}}\right).
\end{equation}
 
Although the accretion rate is defined as the accretion on the cloud and not the sink particle, we assume that both of these accretion rates do not differ significantly and use the accretion rate of the sink particle to calculate the stellar mass.
The estimated masses for sink particles in halo A and B are $\SI{40}{M_\odot}$ and $\SI{100}{M_\odot}$ respectively. For halo C, the most massive clump would correspond to a stellar mass of $\SI{600}{M_\odot}$ and the other two sink particles to approximately $\SI{100}{M_\odot}$ each. No other sink particles are formed.

Note that, according to \cite{Haemmerle18}, the critical accretion rate for the formation of the (high mass) direct collapse black holes could be lower than previously expected. Their calculations suggest an $\dot{\textrm{M}}_\text{crit}$ below $\SI{0.01}{M_\odot {yr}^{-1}}$ rather than the previously suggested $\SI{0.05}{M_\odot {yr}^{-1}}$. In this scenario, the central most massive sink particle for halo C could be a direct collapse black hole.

\subsection{Resolution tests}

During the collapse stage, the resolution is primarily controlled by our Jeans refinement parameter (as opposed to the gas or dark matter Lagrangian refinement criteria).  To explore the impact of varying spatial resolution, we varied the Jeans refinement parameter from 16 to 32 in halo A. The resulting collapse redshift, profile, and overall properties were very similar with the lower-resolution result (although turbulent substructure was better resolved).\footnote{We had to change the starting time of the ionizing radiation to $z=35$ in the higher-resolution case, otherwise, we found that a minihalo collapse was trigged at $z=30$ simply by turning on the radiation.  This occurred because of the (unrealistically) isotropic nature of the uniform background, and was also seen in \cite{Visbal17}. A better fix would be to include non-isotropic radiative transfer, but we found that a simple tweak of the initial redshift avoided this pathological situation.} Because of this similarity, we used Jeans refinement parameter equal to 16 (note that these are both larger than the factor of 4 suggested in the \cite{Truelove97} paper).

We also checked the effect of resolution on the evolution of the sink particle. For halo C, we did a run where we added the sink particle when the simulation reaches 20 levels of refinement, instead of 18 in the default run. (i.e. 4 times better resolution than the default run.) The accretion histories of the sink particles are similar at early times and show a small difference at the late times. However, as the final mass of the Pop III star depends on the accretion rate rate at early times, the properties of the stars and effect of the feedback does not vary much with increased resolution. This is consistent with \citet{Regan18b}, who found little signs of fragmentation for a similar, relatively high, LW radiative background.

\section{Discussion}
\label{sec:discussion}

\subsection{Collapse criterion}
\label{subsec:collapse criterion}

As seen in the previous section, collapse can be delayed to much later times and higher virial mass for strong background ionizing fluxes. 
\cite{Visbal17} also show that for very high background flux, the virial mass at the collapse reaches a saturation value, typically about an order of magnitude higher than the atomic cooling threshold. 
At what background flux is the saturation reached, if reached, depends on the halo properties and history.

For halos A and C, the amount of cold-dense gas and number of clumps at the times of runaway collapse first increases and then decreases with increasing background flux.
For low ionizing flux, increasing the background flux delays the collapse and increases the virial mass and the mass of cold-dense gas at the collapse. 
For high ionizing flux, increasing the background flux does not increase the halo mass at collapse significantly, but decreases the size of the cold neutral core because of stronger ionization.
The number of clumps identified using YT at the runaway collapse also follows a similar pattern.
Thus, for a given halo, a particular value of the background flux gives the highest mass of cold-dense gas and the highest number of clumps at the runaway collapse. 
For halos A and C, we continued the simulation after the runaway collapse for those background fluxes with highest number of fragments. This may represent the most Pop III stars that could form in these halos.

In the case of halo B, it collapses for flux $0.001 F_0$ and $0.01 F_0$. However, for flux equal to or higher than $0.03 F_0$, the gas in the halo does not collapse at all (untill redshift 5). 
The central gas densities in these cases do not seem to increase with time, as it happens in the cases where gas collapses. 
In some cases, densities even continue to decrease. For halo B, we do not get a saturation background flux value, on the other hand the halo does not collapse at all for high background fluxes. 

\begin{figure}
\includegraphics[width=\linewidth]{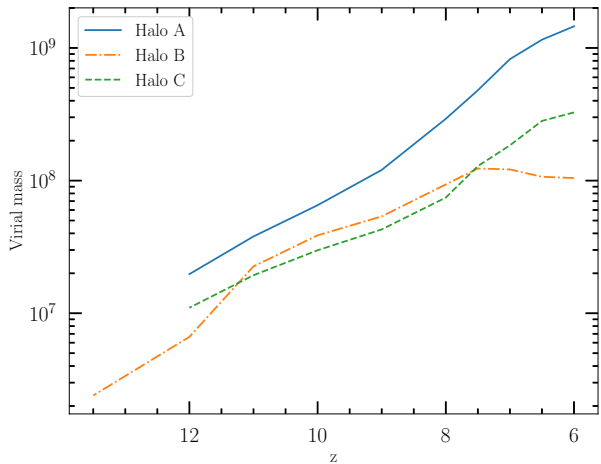}
\caption{Virial mass of the main progenitors of the 3 halos selected with redshift. Halo A and halo C continue to grow untill redshift 6, whereas halo B does not grow after redshift 8. We find that for high ionizing flux halo B does not collapse.}
\label{fig:mvir_vs_z}
\end{figure}

To be able to predict whether a halo with a given background flux will collapse or not, and at what redshift, we investigated the halo merger history. 
We generated a dark matter halo merger tree using \textsc{Rockstar} \citep{Behroozi13_rockstar} and \textsc{consistent-trees} \citep{Behroozi13_trees}. 
Figure \ref{fig:mvir_vs_z} shows the virial mass of main progenitors.
Progenitors for halos A and C continue to grow all the way till redshift 6, whereas halo B does not grow after redshift 8.
Halo B collapses at redshift 8.85 for the background flux of $F = 0.01 F_0$.
For higher ionizing fluxes, halo B does not collapse as it is not able to self-shield because of its stalled growth at low redshift.
From our small of sample of 3 halos, we speculate that this is the most important reason why gas in halo B does not collapse for high background fluxes.
To confirm this hypothesis, we would need do more simulations, particularly ones with stalled halos.

\subsection{Feedback from stars}
Ionizing radiation from the Pop III stars formed would create an HII region around it and also drive an ionization shock.
The growth of the HII region and the shock affects the formation of other Pop III stars. 
We do not include any feedback from the protostar or star in our simulation.
To estimate effect of the feedback from the first star formed on the formation of other stars, we estimate the growth of ionization front and shock associated with it using analytic calculations.
We give an outline of these calculations using halo C as an example, as this is the halo that has the highest clumpiness and multiple sink particles. 

The star corresponding to the central most massive sink particle would have a mass of $\sim \SI{600}{M_{\odot}}$.
According to \cite{Schaerer02}, this star would emit ionizing photons at a rate of $\SI{7.8e50}{s^{-1}}$.
The radial profile of gas density can be well approximated as $n_0 (r/r_c)^{-2}$, consistent with the isothermal density profile, with $n_0 = \SI{1.04e6}{cm^{-3}}$ and $r_c = \SI{0.1}{pc}$ for $r > r_c$ and $n_0$ for $r < r_c$.
We assume that the shock moves out with a velocity of $\sim 10$ km/s.

The HII region will grow as
\begin{equation}
4 \pi  R^2 n(R) dR = \left(\dot{N}_{ion} - 4 \pi \alpha \int_0^Rn^2(r)r^2 dr \right) dt.
\end{equation}
The right hand side denotes the number of available ionizing photons per unit time and $\alpha$ is the recombination rate B coefficient at $T = 10^4$ K. 

Assuming a static density profile mentioned above gives
\begin{equation}
\centering
\frac{dR}{dt} = \frac{\dot{N}_{ion}}{4\pi n_0 r_c^2} - \alpha n_0 r_c \left[\frac{4}{3} - \frac{r_c}{R} \right].
\end{equation}
Initially $dR/dt$ is small, so the evolution of the ionization front is governed by the shock which proceeds as D-type shock just ahead of the ionization front.

As the shock moves forward, the density of the gas in the shocked region decreases. Assuming that the shape of density remains unchanged, the corresponding inner density $n_0$ decreases as $n_0 = \tilde{n}_0 (\tilde{r}_c^2/r_c^2)$, where $\tilde{n}_0$ and $\tilde{r}_c$ are the initial values of $n_0$ and $r_c$.
With these substitutions, $dR/dt$ becomes faster than the shock front at
\begin{equation}
r_c = \frac{4 \pi \alpha \tilde{n}_0^2 \tilde{r}_c^4}{3 \dot{N}}.
\end{equation}
This is where the ionization front starts to lead ahead of the shock and the shock front changes from D-type to R-type. \citep{Franco90, Whalen2004, Kitayama2004, Alvarez2006} For the given parameters, this happens at $r_c$ = 4.7 pc.

For a given density distribution, we can define a critical ionizing photons rate $\dot{N}_\text{crit}$ such that if $\dot{N} > \dot{N}_\text{crit}$, all gas would be ionized, as follows
\begin{equation}
    \dot{N}_{crit} = \frac{16 \pi \alpha \tilde{n_0}^2 \tilde{r_c}^4}{3 r_c}.
\end{equation}
For $\dot{N} = 7.8 \times 10^{50}$ ionizing photons s$^{-1}$, this happens when $r_c \sim \SI{19}{pc}$. When the shock front has reached between $\SI{4.7}{pc}$ and $\SI{19}{pc}$, the ionization front leads ahead of the shock, but only by a small distance.

Until reaching \SI{19}{pc}, the growth of the ionization front is dictated by the growth of the shock. Assuming a shock velocity of \SI{10}{km/s}, the shock would take nearly \SI{2}{Myr} to reach \SI{19}{pc}. After this the ionization front would move very rapidly outwards. This would mean that the formation of two other sink particles at distance $\sim \SI{20}{pc}$, would not be affected by the growth of the HII region from the first star, as they form within 1 Myr after the first star. If the shock speed is as high as \SI{30}{km/s} as \cite{Whalen2004} suggest, the shock would take $\sim 3$ times shorter time.

For halo C, we also see few dense clumps that have not formed a sink particle yet. A clump of diameter 1 pc and density $\SI{e4}{cm^{-3}}$ would have high recombination rate to be ionized completely if it is within ${10-20}$ pc from the center. Such a clump may collapse or be destroyed when the shock front reaches it. If collapse of such clump is triggered, then we may have more star formation fuelled by the radiative feedback from the first star.

The stars formed will also emit LW radiation along with the ionizing radiation. The LW radiation is optically thin and travels close to the speed of light. We estimate the number of LW photons emitted by the first star formed using \cite{Schaerer02} which corresponds to a LW flux of $\sim 10^5 \mathrm{J}_{21}$ at a distance of \SI{20}{pc}. Of the two other sink particles that form at a distance of $\sim \SI{20}{pc}$ from the first sink particle, one forms soon after the first sink, whereas the other one forms nearly 1 Myr later. We expect that the LW radiation from the first star will not affect the evolution of the second star formed, but would alter the evolution of the third sink particle. As $J_{\rm LW}$ of $10^5 J_{21}$ is higher than $J_{\rm crit}$, most of the molecular hydrogen at the location of the third sink particle would be destroyed and the gas temperature would raise to $\sim \SI{e4}{K}$ increasing the accretion rate as well. If the accretion onto the third sink particle continues for long time, it may turn into a DCBH, however it is hard to predict if that would happen here.

\subsection{Observing with JWST}
\cite{JWST} computed that a Pop III galaxy with a stellar mass of $\sim \SI{3e4}{M_{\odot}}$ and few times $\SI{e5}{M_{\odot}}$ could be detected at $z = 6$ with maximal and no nebular emission respectively with an integration time of 100 hours. The limits decrease as $t^{-1/2}_\text{obs}$. This estimate is based on the spectral analysis code \textsc{Yggdrasil} that can identify Pop III galaxies using photometry based on the presence of strong hydrogen and helium lines and the absence of strong metal lines (e.g. \textsc{O ii}, \textsc{O iii}, \textsc{S iii}).

Here we find that for the halos we simulated, even ones with large fragmentation do not produce enough stars to be observed in this limit. For halo C, we estimate a total stellar mass of $\sim \SI{e3}{M_{\odot}}$ which is smaller than the detection limit by an order of magnitude.  \cite{Visbal17} estimate that if all star formation in suppressed in halos up to $10^9 M_{\odot}$, we would get a comoving number density of $10^{-7} \text{Mpc}^{-3}$ which corresponds to 0.002 per JWST field of view per unit redshift at $z = 6$. Thus, both in terms of brightness and their abundances, Pop III galaxies would be unlikely to be detected in an ultra deep field with JWST. A better strategy would be to look near massive galaxies which would be the source for the ionizing photons to create Pop III galaxies in lensed fields. Another possible strategy would be to look for Pop III galaxies forming in atomic cooling halos exposed to just LW photons at lower redshifts as they would have higher abundances.  

\section{Summary and Conclusion}
\label{sec:conclusion}
We performed zoom-in simulations around 3 different halos for various background ionizing fluxes with runaway collapse at $z = 6-15$. Our main conclusions can be summarized as follows.
\begin{enumerate}
    \item Pop III star formation can be suppressed to much lower redshifts in the presence of ionizing radiation from a nearby galaxy. The collapse finally occurs when the gas is able to self-shield itself from the background radiation. 
    \item For very high background ionizing flux, the virial mass at collapse can be up to an order of magnitude higher than the atomic cooling threshold. For a given halo, the amount of cold-dense gas first increases, reaches a maximum value and then decreases with increasing background flux. A similar pattern can be seen in the number of clumps formed at the runaway collapse. The location of the peak would depend on the halo history. For a halo that does not grow at all over a large redshift range, increasing the background flux may photo-evaporate the gas and prevent collapse.
    \item We use YT's clump finder and a sink particle implementation in \textsc{Enzo} to study fragmentation in these Pop III galaxies. Based on their accretion rates, the estimated masses of the Pop III stars range from $40-600$ M$_\odot$.
    \item The most interesting point about these systems is their possible observability with upcoming telescopes such as the James Webb Space Telescope. Halos with pristine gas that are exposed to ionizing radiation would be some of the most massive and late forming galaxies with Pop III stars. In our simulations, we form a handful of Pop III stars in the first 2 Myr. The star formation transitions from Pop III to Pop II when a star explodes as a type II supernova ($M_* < \SI{40}{M_\odot}$) or as a pair-instability supernova ($\SI{140}{M_\odot} < M_* < \SI{250}{M_\odot}$) \citep{Greif15}. With low star formation efficiency, the most massive galaxy in our simulations has $\sim \SI{e3}{M_\odot}$ in stellar mass at $z = 6.54$ which is an order of magnitude lower than the detection limits for an ultra-deep observation using JWST. We therefore conclude that it would be very unlikely to detect such massive late-forming Pop III galaxies with JWST ultra-deep observations and a better strategy could be to look for them in the lensed fields.
\end{enumerate}

\section*{Acknowledgements}
We thank Zoltan Haiman for useful discussions. The computations in this paper were carried out on the NASA Pleiades supercomputer through the NSF supported XSEDE program. The Flatiron Institute is supported by the Simons Foundation.  We also acknowledge support from NSF grants AST-1615955 and OAC-1835509 and NASA grant NNX15AB19G.






\bibliographystyle{aasjournal}
\bibliography{mihir_pop3} 





\end{document}